# Dust in the Cores of Early-Type Galaxies

P.G. van Dokkum

and

M. Franx

Kapteyn Astronomical Institute, P.O. Box 800, NL-9700 AV Groningen, The Netherlands

**ABSTRACT**

The dust properties of all early-type galaxies imaged with the Planetary Camera of the *Hubble Space Telescope* in cycles I – III are examined. Dust is detected in 31 out of 64 galaxies, although the sample is biased against the detection of dust. From the distribution of observed axis ratios of the dust features it is inferred that $78 \pm 16\,\%$ of early-type galaxies contain nuclear dust. The detection rate in radio galaxies is higher ($72 \pm 16\,\%$) than in radio-quiet galaxies ($33 \pm 9\,\%$). In radio galaxies with double radio structure, the dust is usually found perpendicular to the radio axis. Dust masses range from $10^3 - 10^7$ solar masses. From the dependence of the misalignment angle (the angle between dust major axis and galaxy major axis) on size and appearance of the dust, it is found that dust features with semi-major axis size $a_\mathrm{d} > 250\,\mathrm{pc}$ are generally not settled. If it is assumed that dust lanes with $a_\mathrm{d} < 250\,\mathrm{pc}$ are in equilibrium in the galaxy potential, the observed distribution of misalignment angles can only be explained if elliptical galaxies have a mean triaxiality $\langle T \rangle \geq 0.4$. The best fit is obtained for $\langle T \rangle = 0.8$, higher than values derived through other methods. This difference cannot be explained by a dependence of $\langle T \rangle$ on radius. Therefore, it seems dust is not always completely relaxed, even on the smallest scales. Furthermore, the rotation axis of the dust does not generally coincide with that of the stars. This seems to suggest that the dust is acquired from outside, and has some way of losing its angular momentum efficiently.

*Subject headings:* galaxies: elliptical, galaxies: nuclei, galaxies: structure of, galaxies: ISM

## 1. Introduction

It has become clear that many elliptical galaxies contain a significant amount of interstellar medium (ISM) (e.g., Roberts *et al.* 1991), primarily in the form of hot, X-ray emitting gas, with masses $\sim 10^{10}\,M_\odot$ (cf. Table 3, Col. 5). A small fraction is in the form of dust, and associated cold gas (Kormendy & Djorgovski 1989; Forbes 1991; Goudfrooij 1994). Although the mass of the cold component of the ISM is small ($\sim 10^7\,M_\odot$), the study of dust in early-type galaxies is of interest, in several respects.

The origin of the dust is related to the evolutionary history of the parent galaxies. In a systematic study, it may be possible to discriminate between an internal origin (stellar mass loss; Knapp *et al.* 1992), or an external origin (accretion; Forbes 1991). Closely related to the origin of the dust is its dynamical state. Is it in equilibrium in the galaxy potential, or is it not (completely) settled? This question is linked to the intrinsic shape of elliptical galaxies: if the dust is settled, the dust lane indicates a plane in the galaxy in which stable closed orbits are allowed (van Albada *et al.* 1982; Lake & Norman 1983; Merritt & de Zeeuw 1983). The fact that ellipticals may be triaxial is a serious complication, which was usually not considered in previous studies (e.g., Bertola & Galletta 1978, Hawarden *et al.* 1981, Ebneter & Balick 1985, Zeilinger *et al.* 1990). Finally, correlations between dust content and other observables have been suggested. Dust may correlate with radio properties (e.g., Kotanyi & Ekers 1979, Marston 1988, Scoville *et al.* 1991), indicative of a relation with active galactic nuclei (AGN), and with the hot gas component of the ISM, providing information regarding the origin and fate of the dust (Gunn 1979).

There are obvious advantages when observing dust at small ($\lesssim 1\,\mathrm{kpc}$) scales. In the nuclear regions of a galaxy, settling timescales for dust are short ($\sim 10^8\,\mathrm{yr}$), which is important in view of the dynamical state of the dust, and its relation to the intrinsic shape of the parent galaxy. Furthermore, any correlation with AGN phenomena is expected to be stronger than for large scale dust, since the distance to the active nucleus is small. Nuclear dust has been detected in several galaxies (e.g., Möllenhoff & Bender 1987, Kormendy & Stauffer 1987). Resolution is generally the limiting factor in detecting nuclear dust. Therefore, it was not until the launch of the *Hubble Space Telescope* (HST) that it became clear that in fact many early-type galaxies have dust in the central regions (Forbes *et al.* 1994; Kormendy *et al.* 1994; Jaffe *et al.* 1993, 1994).

From the launch of the HST until its repair in December 1993 82 early-type galaxies were imaged with the Planetary Camera (PC), by various observers and for various projects. By now, all these data are in the public domain and, although far from a complete sample, this data set offers the oppertunity to study characteristics of the cores of elliptical galaxies in a systematic way, with unprecedented resolution. The galaxies have been studied extensively in the framework of the projects for which they were observed. Here, we examine *all* early-type galaxies imaged with the HST in cycles I – III, to answer the specific questions related to the frequency and structure of nuclear dust, and to learn the dynamical state of the dust, in relation to the intrinsic shapes of the host galaxies.

## 2. Sample selection

We only considered galaxies observed with the Planetary Camera of the HST in cycles I – III, with de Vaucouleurs type < 0. Of these 82 galaxies, the data were requested from the Space Telescope Science Institute (ST ScI) archive. In the case of images in several colors, we took the F555W image (roughly corresponding to the Johnson $V$ band), or the image in the color closest to it. The data of one galaxy (NGC 855) were not public at the time of request, two probably suffer from pointing errors (NGC 201 and NGC 1705), and two are severely saturated (NGC 4151 and NGC 5548). The detection of faint dust features depends critically on the S/N ratio of the image. We applied a S/N criterion, in order to obtain a more homogeneous sample:

$$m_V - 2.5 \log t \leq 6, \qquad (1)$$

with $t$ the exposure time in seconds, thus reducing the sample by 13 galaxies: NGC 1331, NGC 1409, NGC 1427, NGC 2636, NGC 4458, NGC 4467, NGC 4486B, VCC 1199, VCC 1440, VCC 1545, VCC 1627, UGC 1597 and UGC 2456. In Table 2 the remaining 64 galaxies are listed.

Selection effects may play an important role in this kind of work. In Table 1 we list the descriptions of the projects our sample is drawn from. The Faber sample is 'heavily weighted in favor of objects that did not show dust in ground-based images' (Kormendy *et al.* 1994). The description of the 'nearly normal' galaxies of the Westphal project in the proposal abstract is: 'The sample of objects contains several normal ellipticals covering a broad range in nuclear surface brightness and concentration class, several nearby galaxies covering a range of Hubble types, and a few Seyfert and otherwise slightly abnormal nuclei.' Consequently, there is a bias against the detection of dust, and a bias in favor of kinematically distinct cores.

## 3. Observations

We obtained all data from the ST ScI archive, operated by the Association of Universities for Research in Astronomy, Inc., for the National Aeronautics and Space Administration. All observations were made with the Wide Field and Planetary Camera (WFPC), in Planetary Camera mode (f/30). The field-of-view is 66 × 66 arcsec, and the pixel size 0.043 arcsec. The images were taken through the F555W filter, except NGC 2110, which is observed using the F492M filter.

We estimate the resolution of the deconvolved images is $\lesssim 0.''1$.

## 4. Reduction

For the reduction we primarily used the Space Telescope Science Data Analysis Software (STSDAS), implemented in the IRAF package. We did not recalibrate any data; for each file it was checked whether the right calibration files were used. According to Baum (1993), differences after a recalibration will generally be small. Cosmic rays were removed using the COMBINE task. In the case of single images, cosmic rays were identified in a manner similar to the identification of cosmic rays and bad pixels remaining after combining cosmic ray split images. Delta flats, obtained from ST ScI, were compared to the residual of a harmonics fit to the galaxy, and from this a smoothed mask was constructed to replace the measles with the values from the harmonics fit. Bad columns and left over bad pixels were identified in the residual of the fit as many $\sigma$ deviations and then replaced by the fit values. For the harmonics fit the GALPHOT package (Franx *et al.* 1989a; Jørgensen *et al.* 1992) was used.

We used Tiny Tim 2.1 to simulate the Point Spread Function (PSF) of the HST optics. Essentially the only free parameter is the pointing jitter. Before *Space Telescope* was launched, the expected jitter was 7 mas in fine lock and 23 mas in coarse lock. However, the amplitude of the jitter is unpredictable and varies significantly, especially in coarse lock (Faber 1990; Burrows *et al.* 1991). From a star imaged in coarse lock the jitter was estimated to be $30 \pm 5$ mas. In this work, a value of 7 mas is used if fine tracking was utilized and 25 mas for images observed in coarse lock. For the deconvolution, we used the STSDAS implementation of the Lucy-Richardson algorithm (Richardson 1972; Lucy 1974; Baade & Lucy 1990), typically with 80 iterations. Lucy's method is not suitable for the deconvolution of point sources. When present, the pointsource was removed by subtracting a scaled and subsampled PSF from the image (as described in Forbes *et al.* 1994) before deconvolution with the Lucy algorithm.

The HST data of some galaxies in our sample are described in the literature. We have made use of this by comparing the deconvolved images with the published images. No significant differences were found. At the time of this project, data are published of the Faber project (Kormendy *et al.* 1994; preliminary report), the Jaffe project (Jaffe *et al.* 1994; van den Bosch *et al.* 1994; Ferrarese *et al.* 1994), and some individual galaxies: NGC 1275 (Holtzman *et al.* 1992), NGC 1399 (Stiavelli *et al.* 1993), NGC 4261 (Jaffe *et al.* 1993), NGC 4486 (Lauer *et al.* 1992), IC 1459 (Forbes *et al.* 1994), NGC 7252 (Whitmore *et al.* 1993), and NGC 7457 (Lauer *et al.* 1991).

## 5. Analysis

The detection of faint dust features is facilitated when the smooth light of the galaxy is subtracted from the image. We fitted ellipses with the GALPHOT package. If dust features were detected after a first fit a second fit was made, with the dust masked. When the first fit was severely affected by dust in the inner regions, we also fixed the position angle and / or the ellipticity of the ellipses for these parts in the second fit.

The dust features were classified by their morphology. We distinguished between regular and irregular (patchy), and (mirror) symmetric and asymmetric with respect to the center of the galaxy. Where possible, we estimated the position angle $PA_d$ of the longest axis of the dust feature, and the axis ratio $\frac{b_d}{a_d}$, where $a_d$ is the semi-major axis, and $b_d$ is the semi-minor axis of the dust feature. These quantities can have considerable uncertainty. The misalignment angle $\Psi$ is the angle between the longest axis of the dust feature and the galaxy major axis: $\Psi = \mathrm{PA}_{\mathrm{gal}} - \mathrm{PA}_{\mathrm{d}}$; $-90° < \Psi < 90°$. The derived parameters of the dust features and of the ellipse fits to the galaxies are listed in Table 2.

For each galaxy, an absorption map was created by deviding the deconvolved image by a model, constructed from the ellipse fit. For each dust feature in these maps, the dust mass was estimated using $M_{\mathrm{dust}} = \Sigma \langle A_V \rangle \Gamma_V^{-1}$, with $\Sigma$ the area of the feature, $\langle A_V \rangle$ the mean absorption in that area and $\Gamma_V$ the visual mass absorption coefficient (Sadler & Gerhard 1985). The total dust mass in a galaxy is the sum of the estimated masses of all dust features in that galaxy. Differences between Galactic and elliptical extinction curves are generally small (Goudfrooij *et al.* 1994). Therefore, we adopt the Galactic value $\Gamma_V \sim 6 \times 10^{-6} \, \mathrm{mag \, kpc^2} \, M_\odot^{-1}$. Galactic gas to dust ratios ($M_{\mathrm{gas}}/M_{\mathrm{dust}} \sim 1.3 \times 10^2$) are assumed; hereafter, dust mass $M_{\mathrm{d}}$ refers to $M_{\mathrm{gas}} + M_{\mathrm{dust}}$. The derived dust masses are listed in Table 2. The typical uncertainty is at least 50 %. Adopted distances are listed in Table 3, along with other basic parameters of the galaxies studied. A Hubble constant $H_0 = 50 \, \mathrm{km \, s^{-1} \, Mpc^{-1}}$ is assumed throughout the paper.

A mosaic of the galaxy cores that clearly show evidence for dust absorption is presented in Fig. 8. The images were divided by a model, constructed from the ellipse fit. Consequently, the images show the fraction of absorption of the galaxy light. The scale of each image is 128 × 128 pixels, or $5\rlap{.}''5 \times 5\rlap{.}''5$. The ellipse indicates the average position angle and ellipticity of the ellipse fit values between $10''$ and $15''$. The line in the lower right corner of each panel shows the estimated position angle of the dust. The center of each galaxy is at the center of the corresponding image. The image of NGC 5102 shows residuals of the saturated pointsource present in the galaxy image. The streak of dust in this galaxy is to the left of the image, a patch to the right.

## 6. Properties of nuclear dust

### 6.1. Frequency of nuclear dust in early-type galaxies

In our sample, 31 out of 64 galaxies (48 %) show nuclear dust absorption. Most galaxies have dust axis ratios $\frac{b_d}{a_d} < 0.5$, which suggests that face-on dust features had a low probability to be detected. This is consistent with the fact that the local absorption is lower than 0.2 magnitudes for all dust features with axis ratios $\frac{b_d}{a_d} > 0.5$.

In Fig. 1 the cumulative distribution of observed axis ratios $\frac{b_d}{a_d}$ is shown. The expected distribution for dust on circular orbits with zero intrinsic thickness is a straight line. Assuming completeness for $\frac{b_d}{a_d} < 0.5$, the intercept of the solid line with the vertical axis shows the number of dusty cores, corrected for the selection effect. For 4 galaxies no axis ratio could be assigned to the dust, which brings the total estimate to $50 \pm 10$ out of 64, or $78 \pm 16$ %. The dotted line in Fig.

1 shows the expected cumulative distribution for dust on elliptical orbits, with intrinsic axis ratio $\left(\frac{b_d}{a_d}\right)_i = 0.8$. Our estimate does not depend on the assumption that the dust is on circular orbits.

### 6.2. Nuclear dust and hot gas

Many ellipticals are embedded in a hot, X-ray emitting gas (e.g., Roberts *et al.* 1991). The effect on dust is probably complex: friction with the gas may cause dust to lose angular momentum and fall to the nucleus (Gunn 1979), and dust grains can be destroyed in interactions with the gas (Goudfrooij 1994). In Fig. 2 the nuclear dust mass is plotted against hot gas mass (a), taken from Roberts *et al.* (1991), and absolute magnitude (b). No relations can be inferred from Fig. 2. Goudfrooij (1994) finds an anti-correlation between dust mass and X-ray mass, and no dependence of dust mass on absolute magnitude. We confirm the latter statement.

### 6.3. Size and regularity: the 'relaxation radius'

Since the relaxation time for dust is dependent on the dynamical time, the fraction of relaxed dust lanes might be expected to increase with decreasing $r$. The correlation between spatial scale and dynamical state of the dust can be investigated by plotting $|\Psi|$, the angle between major axis of the galaxy and major axis of the dust feature, against the semi-major axis size $a_d$ (Fig. 3). To first order, $|\Psi|$ is a measure of the dynamical state of the dust: for dust which is on closed orbits in the galaxy potential $|\Psi|$ is, on average, close to 0° or 90° (cf. Sect. 7.).

Dust lanes with $a_d > 250\,\mathrm{pc}$ have a uniform distribution of $|\Psi|$ and usually an irregular appearance. Apparently, dust is generally not settled at scales $r > r_\mathrm{rel}$, where $r_\mathrm{rel} \sim 250\,\mathrm{pc}$ is the 'relaxation radius'. The dynamical state of dust with sizes $a_d < r_\mathrm{rel}$ can only be discussed in relation to the intrinsic shape of the underlying galaxies (Sect. 7.).

### 7. The dynamical state of dust on scales $< r_\mathrm{rel}$

We test the hypothesis that all dust lanes smaller than 250 pc are in equilibrium. The observed distribution of $|\Psi|$ for these dust lanes is modeled assuming different galaxy intrinsic shapes.

The orientation of a relaxed dust lane in a galaxy indicates the orientation of a plane in which stable closed orbits are allowed. In oblate galaxies this is the plane perpendicular to the short axis, in prolate galaxies the plane perpendicular to the long axis. Assuming galaxies are either oblate or prolate, galaxy shapes have thus been derived from large scale dust lanes by e.g., Bertola & Galletta (1978), Hawarden *et al.* (1981). In a triaxial galaxy, stable closed orbits are allowed in two planes: the plane containing the short and the intermediate axis, and the plane containing the long and the intermediate axis. The projections of these planes are generally not parallel to either the observed major axis or the observed minor axis. The misalignment angle $|\Psi|$, defined as the angle $0 \leq |\Psi| \leq 90$ between the major axis and the dust lane, is determined by the shape of the galaxy, the plane the dust is in, and the viewing angle.

The shape of the galaxy can be parametrized with the triaxiality parameter $T$, defined as

$$T = \frac{a^2 - b^2}{a^2 - c^2}, \qquad (2)$$

with $a$ the long axis, $b$ the intermediate axis and $c$ the short axis. Oblate galaxies have $T = 0$ and prolate galaxies $T = 1$. Franx *et al.* (1991) (FIZ91) showed that, at fixed viewing angles, $|\Psi|$ is equal for galaxies with equal $T$.

We first assume all dust lanes are in the plane perpendicular to the short axis, since there are no dust lanes in our subsample with $|\Psi| > 50°$. Because the viewing angles for individual galaxies are unknown, we apply the method used by FIZ91. They derive probability distributions of $\Psi_{\rm kin}$, where $\Psi_{\rm kin}$ is the angle between minor axis and kinematic axis of the stars, for different values of $T$, by integrating over all possible viewing angles.

If the dust is on circular orbits, the major axis of the dust is perpendicular to the intrinsic short axis for all viewing angles. The maximaly triaxial models from FIZ91 imply a mean intrinsic axis ratio for the dust disk of $\left(\frac{b_{\rm d}}{a_{\rm d}}\right)_{\rm i} \approx 0.96$. This means we can safely assume the dust is on circular orbits. Therefore, the probability distribution of $|\Psi|$ is equal to the probability distribution of the angle between the intrinsic short axis and the observed minor axis. The probability distribution of $|\Psi|$ is equal to the probability distribution of $|\Psi_{\rm kin}|$, if the intrinsic misalignment $\phi_{\rm int} = 0$ (see FIZ91).

In Sect. 6.1., it was inferred that dust lanes with inclinations $< 60°$ are usually not detected. This implies a strong bias towards smaller values of $|\Psi|$ (see FIZ91). We used Monte-Carlo simulations to derive the probability distribution of $|\Psi|$ for different values of $T$, only allowing viewing angles for which $\frac{b_{\rm d}}{a_{\rm d}} < 0.5$. In the simulations, we incorporated a measurement uncertainty of $5°$, which is the average error in $\Psi$ for dust lanes with $a_{\rm d} < r_{\rm rel}$.

In Fig. 4 (a), the observed distribution of $|\Psi|$ is shown, for the subsample of dust lanes with $a_{\rm d} < r_{\rm rel}$. The histogram in Fig. 4 (b) shows the cumulative distribution. The lines represent the derived probability distributions for different values of $T$. According to a Kolgomorov-Smirnov (KS) test, an average triaxiality $\langle T \rangle < 0.4$ can be ruled out at the 95 % confidence level. The best fitting curve is for $\langle T \rangle = 0.8$.

This high value for the triaxiality is inconsistent with the results of FIZ91, who find a mean triaxiality of $\langle T \rangle \leq 0.4$ from observed minor axis rotation. We consider the possibility that this difference is caused by a dependence of $T$ on radius, i.e., that galaxies are close to oblate at large radii, but have highly prolate nuclei.

Such a change in triaxiality causes position angle twists in the projected stellar body. These twists are a function of the change in triaxiality $\Delta T$, and the viewing angle. By integrating over all possible viewing angles, we can predict the observed position angle twists, and compare them with the data. In Fig. 5, we have plotted the measured position angle twist between $r = 25''$ and $r = 1''\!.5$ as a function of ellipticity at $r = 25''$. The dashed line represents the expected $1\sigma$ spread, calculated with Monte-Carlo simulations, for a triaxiality change of 0.4. The data are inconsistent with a $\Delta T$ close to 0.4. A Student $t$ test gives an upper limit for $\Delta T$ of 0.20 at the 95 % confidence level.

## 8. Dust rotation in relation to stellar kinematics

For 11 galaxies in the sample the kinematic axis of the stars is known. Kinematic misalignment angles $\Psi_{\rm kin}$ are taken from Davies & Birkinshaw (1988), Franx *et al.* (1989b) and Jedrzejewski & Schechter (1989). In Fig. 6 $\Psi_{\rm kin}$ is plotted against $\Psi$. No galaxies with misalignments significantly different from zero are located on the line $\Psi_{\rm kin} = \Psi$.

Galaxies are expected to have $\Psi = \Psi_{\rm kin}$ if dust lanes are relaxed and the intrinsic misalignment between kinematic axis and shortest axis $\phi_{\rm int} = 0$. Apparently, the dust and the stars are generally not kinematically coupled.

## 9. Nuclear dust and radio properties

It has been suggested that galaxies with dust lanes more often have a radio detection than dustless galaxies (Sadler & Gerhard 1985; Marston 1988). According to Gregorini *et al.* (1989) radio structures in dusty early-type galaxies are small and weak. Véron-Cetty & Véron (1988) claim there is *no* correlation between radio detection and dust content, for a complete sample of 78 early-type galaxies.

Scattered through the literature, we found a detection at 6 cm for 25 galaxies in our sample of 64. The detection rate of dust in radio galaxies is 18/25 ($72 \pm 16$ %), compared to 13/39 ($33 \pm 9$ %) for radio quiet galaxies. We note, however, that our sample is far from complete.

Kotanyi & Ekers (1979) (KE79) found that large scale dust lanes are usually perpendicular to the radio axis in active galaxies. Remarkably, the radio axis seems to be oriented randomly with respect to the galaxy major axis (Sansom *et al.* 1987). The relation between radio axis and galaxy major axis changes with redshift: at large $z$, optical emission seems to be aligned with radio emission (e.g., Eales 1992).

Any correlation between radio structure and dust orientation is likely to be stronger near the nucleus, since there timescales are short and the distance to the active nucleus small. Still, for dust lanes at HST resolution $r/r_{\rm g} \sim 10^7$, where $r_{\rm g}$ is the gravitational radius of a massive black hole.

Also, Lense-Thirring precession (Bardeen & Petterson 1975) of the inertia frame around the black hole might make the direction of outflow to be rather insensitive of the direction of the angular momentum vector of the infalling material (Rees 1978; Rees 1984). Therefore, no correlations between dust lane and radio axis orientation may be required from a dynamical point of view.

In our sample 17 galaxies show extended radio structure, including sources with only an extension on mas scales or just lobes, with no jet detected. Therefore, not all of the 17 galaxies have a jet according to the Bridle & Perley (1984) definition. In Table 4 the radio morphologies of this subsample are described. In the following, we use the arcsec scale position angle. The galaxy NGC 4278 is not considered, since the large scale radio emission originates from an extended H I cloud (Raimond *et al.* 1981).

We define $|\Psi_{\text{stars}-\text{radio}}|$ as the angle $0 \leq |\Psi_{\text{stars}-\text{radio}}| \leq 90$ between radio axis and major axis of the galaxy, and $|\Psi_{\text{radio}-\text{dust}}|$ as the angle $0 \leq |\Psi_{\text{radio}-\text{dust}}| \leq 90$ between the dust lane and the radio axis. KE79 find $|\Psi_{\text{radio}-\text{dust}}| = 90° \pm 30°$. Figure 7 shows the distributions of $|\Psi_{\text{stars}-\text{radio}}|$, $|\Psi|$ and $|\Psi_{\text{radio}-\text{dust}}|$ for the subsample of double radio galaxies. According to a KS test a uniform distribution of $|\Psi_{\text{radio}-\text{dust}}|$ can be ruled out at the 95 % confidence level. Therefore, we confirm the conclusion of KE79 that $\Psi_{\text{radio}-\text{dust}}$ peaks near 90°.

## 10. Discussion and conclusions

From the analysis of Sect. 6.3. and Sect. 7., two points can be made: (1) dust is generally not settled at scales $r \gtrsim 250$ pc and (2) at smaller scales dust is not always completely settled. As part of the analysis, we have shown that elliptical galaxies do not have morphologically distinct nuclei, i.e., there is no large difference in triaxiality between outer parts and inner (arcsec-scale) parts. The maximum difference in triaxiality between $r = 25''$ and $r = 1''\!.5$ allowed by the data is $\Delta T = 0.20$.

The fact that most dust features are not in equilibrium in the galaxy potential may influence the estimated frequency of nuclear dust (cf. Sect. 6.1): an explanation for the apparent deficiency of face-on dust is that the morphology is always quite irregular, so that there is always some edge-on system present, which would be dominating. However, the fact that regular and symmetric dust features also show a deficiency of face-on systems, strongly suggests that we *do* miss a considerable fraction of dust features due to their orientation. Furthermore, any estimate from visual absorption alone should be regarded as a lower limit: it may well be that color images, made with the refurbished HST, will make it possible to detect dust in an even larger fraction of early-type galaxies.

Interestingly, two of the dust lanes with $a_{\text{d}} < 250$ pc are warped. Both are in galaxies with kinematically distinct cores (IC 1459 and NGC 7626). Balcells & Carter (1993) deduce the core of the latter galaxy is a young ($\leq 10^8$ yr) merger remnant. Furthermore, in Sect. 7. only intrinsic short axis rotation was considered. It may be that in NGC 3379, which has $|\Psi| = 48° \pm 3°$, the dust rotates around the intrinsic long axis, and the remaining 6 galaxies have completely relaxed dust, orbiting around the intrinsic short axis of close to oblate galaxies. Of course, no statistically significant statements can be made regarding the intrinsic shape of ellipticals when only these galaxies are considered.

The conclusions of Sect. 6.3 and Sect. 7 may hold clues for the origin of the dust: is it captured or does it originate from stellar mass loss? Detailed model predictions for the settling timescale and appearance of dust do not exist, either for capture or stellar mass loss. Presumably, the dynamics of the hot X-ray gas also play a role here. However, it is hard to reconcile the random distribution of dust lane orientations (Fig. 3) with an internal origin. This is demonstrated in Fig. 6: a comparison of dust rotation axis and stellar rotation axis shows that the angular momentum of many of the dust systems is very different from that of the stars. Four galaxies (out of 11) in Fig. 6 have both $\Psi = 0$ and $\Psi_{\rm kin} = 0$, consistent with an internal origin. However, it would be surprising if in *all* galaxies the captured material is still in a non-equilibrium state. An interesting test would be to investigate whether the stars and the gas in these galaxies have the same sense of rotation. The misalignment of stellar rotation axes and dust rotation axes is all the more remarkable, since the stars are expected to have a present day mass loss rate of $\sim 0.1 - 1 M_\odot \, {\rm yr}^{-1}$, and produce a considerable amount of cold gas, and associated dust (e.g., Goudfrooij *et al.* 1994). Apparently, we are not seeing this gas.

Since the estimated frequency of nuclear dust is $78 \pm 16\,\%$, an external origin of the dust implies most elliptical galaxies have had some form of tidal interaction in the past. The settling time for dust at $r_{\rm rel}$ is $\sim 10^8\,{\rm yr}$; therefore, the dust seems to be acquired relatively recently. We note, however, that detailed predictions for the settling time of a dust lane in the presence of hot X-ray gas do not exist at present.

Important in view of both the origin of the dust and its dynamical state is the relation of the dust with radio properties. Although some unknown selection effect may be involved, we find the dust detection rate is higher in radio galaxies ($72 \pm 16\,\%$) than in radio-quiet galaxies ($33 \pm 9\,\%$), suggesting a connection between the accretion of the cold material and the nuclear activity.

Also, we find the dust lane is usually perpendicular to the radio jet ($|\Psi_{\rm radio-dust}|$ peaks near $90°$), confirming the KE79 result. This has the consequence that either the radio axis is generally aligned with the minor axis for the galaxies in our sample, or that the dust somehow aligns with the central engine, irrespective of the galaxy potential, thus yielding uniform distributions of both $|\Psi_{\rm stars-radio}|$ and $|\Psi|$. Judging from Fig. 7, there seems to be a trend towards larger angles in the distribution of $|\Psi_{\rm stars-radio}|$ *and* a trend towards a more uniform distribution of $|\Psi|$ than for the whole sample, but both effects are not statistically significant.

In conclusion, dust is a very common feature in elliptical galaxies, it seems to be acquired relatively recently and it has some way of losing its angular momentum efficiently.

We thank Tim de Zeeuw for a thorough reading of the manuscript, and the anonymous referee for many helpful suggestions.

Table captions:

TABLE 1. Projects of the Principal Investigators (PIs).
TABLE 2. Nuclear Dust Morphologies of 64 Bright Early-Type Galaxies.
TABLE 3. Basic Data and X-ray Masses of Sample Members.
TABLE 4. Sample Members with Extended Radio Structure.

Description of Table 2:

Col. (3) Appearance of the dust feature: R regular, I irregular or patchy, S mirror symmetric and A asymmetric with respect to the center of the galaxy.
Col. (4) Position angle of the major axis of the dust feature, measured North through East.
Col. (5) Axis ratio $\frac{b_d}{a_d}$, where $a_d$ is the semi-major axis and $b_d$ is the semi-minor axis of the dust feature.
Col. (6) Position angle of the major axis of the galaxy, as determined from the average position angle from the ellipse fit, between $10''$ and $15''$ from the center of the galaxy.
Col. (7) Ellipticity of the galaxy $\epsilon = 1 - \frac{b_{gal}}{a_{gal}}$, where $a_{gal}$ is the major axis and $b_{gal}$ is the minor axis of the galaxy, determined analogous to the position angle of the galaxy (Col. 6).
Col. (8) Misalignment angle between major axis of the galaxy and major axis of the dust feature, defined as $\Psi = PA_{gal} - PA_d$; $-90° < \Psi < 90°$.
Col. (9) Semi-major axis size of the dust feature. Adopted distances are listed in Table 3, Col. 3. ($H_0 = 50\,\mathrm{km\,s^{-1}\,Mpc^{-1}}$)
Col. (10) $M_d$ refers to $M_{dust} + M_{gas}$, assuming $M_{gas}/M_{dust} \sim 1.3 \times 10^2$.

Notes to Table 2:

[a] Dark features can be explained by absorption in the high velocity system, a population effect, or dust absorption in the galaxy (Holtzmann *et al.*, 1992, AJ 103, 691).
[b] Ellipse fit not possible, due to the irregular structure of the galaxy.
[c] From Dutch Telescope 0.9 m image, unpublished.
[d] Sage & Galletta, 1993, ApJ 419, 544.
[e] Indication of stellar ring of $1''2$ at position angle $40°$.

Description of Table 3:

Col. (2) Galaxy type from de Vaucouleurs *et al.* (1991).
Col. (3) Distances taken from Faber *et al.* (1989), Tully (1988), or Huchra *et al.* (1983), assuming $H_0 = 50\,\mathrm{km\,s^{-1}\,Mpc^{-1}}$.

Col. (5) From Roberts *et al.* (1991).

Notes to Table 3:

[a] Galaxy not listed in Roberts *et al.* (1991).

References for Table 4:

Table 1. Projects of the principal investigators (PIs).

| PI | Title | # gal |
|---|---|---|
| S. Faber | Cores of early-type galaxies | 28 |
| J. Westphal | Nuclei of nearly normal galaxies | 12 |
| W. Jaffe | Black holes, stellar dynamics and populations in the nuclei of a complete sample of elliptical galaxies | 11 |
| G. Illingworth | Ellipticals with kinematically-distinct nuclei | 6 |
| J. Westphal | Galaxies and clusters | 2 |
| J. Westphal | Compact blue objects in NGC 1275 | 1 |
| J. Westphal | Peculiar and interacting galaxies | 1 |
| J. Westphal | SAT observation: nucleus of a nearby normal galaxy | 1 |
| B. Whitmore | High-resolution imaging of colliding and merging galaxies | 1 |
| A. Wilson | Ionizing cones, obscuring tori and the narrow line regions of Seyfert galaxies | 1 |

Table 2. Nuclear dust morphologies of 64 bright early-type galaxies.

| Ident | Dust structure | App | $PA_d$ | $\frac{b_d}{a_d}$ | $PA_{gal}$ | $\epsilon$ | $\Psi$ | $a_d$ | $\log M_d$ |
|---|---|---|---|---|---|---|---|---|---|
| | | | (°) | | (°) | | (°) | (kpc) | ($M_\odot$) |
| (1) | (2) | (3) | (4) | (5) | (6) | (7) | (8) | (9) | (10) |
| N221 | ? | ... | ... | ... | 161 | 0.29 | ... | ... | ... |
| N524 | Tight spiral pattern | RA | ... | 0.9 | 50 | 0.03 | ... | 1.99 | 6.1 |
| N596 | ... | ... | ... | ... | 65 | 0.06 | ... | ... | ... |
| N720 | ... | ... | ... | ... | 140 | 0.15 | ... | ... | ... |
| N1023 | ? | ... | ... | ... | 84 | 0.25 | ... | ... | ... |
| N1052 | Lane and wisps | IS | 23 | 0.4 | 113 | 0.32 | 90 | 0.30 | 5.6 |
| N1172 | Lane | IS | 174 | 0.3 | 44 | 0.09 | 50 | 0.97 | 6.3 |
| N1275 | ?[a] | ... | ... | ... | ...[b] | ... | [b] ... | ... | ... |
| N1316 | Complex | IA | ... | ... | 59 | 0.11 | ... | 30[c] | 9.1[d] |
| N1399 | ... | ... | ... | ... | 113 | 0.12 | ... | ... | ... |
| N1400 | ... | ... | ... | ... | 34 | 0.13 | ... | ... | ... |
| N1426 | ... | ... | ... | ... | 104 | 0.42 | ... | ... | ... |
| N1439 | Lane | RS | 37 | 0.3 | 33 | 0.27 | −4 | 0.06 | 4.0 |
| N1600 | ... | ... | ... | ... | 10 | 0.45 | ... | ... | ... |
| N1700 | Warped broad lane | IS | 55 | 0.4 | 90 | 0.31 | 35 | 0.77 | 6.0 |
| N2110 | Dusty spiral disk | RA | ... | 0.7 | 166 | 0.33 | ... | 0.47 | 5.0 |
| N2832 | ... | ... | ... | ... | 156 | 0.32 | ... | ... | ... |
| N3115 | ... | ... | ... | ... | 40 | 0.30 | ... | ... | ... |
| N3311 | Donut-like | IS | 40 | 0.4 | 43 | 0.18 | 3 | 0.53 | 6.2 |
| N3377 | ? | ... | ... | ... | 40 | 0.57 | ... | ... | ... |
| N3379 | Ring | RS | 120 | 0.2 | 72 | 0.08 | −48 | 0.09 | 3.9 |
| N3384 | ... | ... | ... | ... | 47 | 0.36 | ... | ... | ... |
| N3599 | U-shaped lane | IA | ... | 0.6 | 15 | 0.15 | ... | 0.13 | 4.6 |
| N3605 | ... | ... | ... | ... | 20 | 0.47 | ... | ... | ... |
| N3608 | ? | IA | ... | ... | 80 | 0.21 | ... | ... | ... |
| N4150 | Irregular lanes | IA | 140 | 0.3 | 148 | 0.30 | 8 | 0.39 | 5.7 |
| N4168 | ... | ... | ... | ... | 128 | 0.18 | ... | ... | ... |
| N4261 | Donut | RS | 164 | 0.4 | 163 | 0.33 | −1 | 0.17 | 5.4 |
| N4278 | Complex | IS | 86 | 0.2 | 15 | 0.17 | −71 | 0.50 | 5.7 |
| N4365 | ... | ... | ... | ... | 45 | 0.25 | ... | ... | ... |
| N4374 | Multiple lanes | RS | 80 | 0.3 | 129 | 0.21 | 49 | 0.68 | 6.0 |
| N4387 | ... | ... | ... | ... | 147 | 0.28 | ... | ... | ... |
| N4434 | Complex | IA | ... | ... | 25 | 0.08 | ... | 0.20 | 4.3 |
| N4472 | One-sided streak | RA | 334 | 0.2 | 161 | 0.12 | 7 | 0.09 | 3.3 |
| N4473 | ... | ... | ... | ... | 92 | 0.43 | ... | ... | ... |
| N4476 | Spiral-like | RS | 29 | 0.4 | 48 | 0.07 | 19 | 0.81 | 6.3 |
| N4478 | ... | ... | ... | ... | 152 | 0.18 | ... | ... | ... |
| N4486 | ... | ... | ... | ... | 159 | 0.03 | ... | ... | ... |
| N4494 | Ring | RS | 8 | 0.4 | 1 | 0.15 | −7 | 0.07 | 4.2 |

Table 2—Continued

| Ident | Dust structure | App | $PA_d$ (°) | $\frac{b_d}{a_d}$ | $PA_{gal}$ (°) | $\epsilon$ | $\Psi$ (°) | $a_d$ (kpc) | $\log M_d$ ($M_\odot$) |
|---|---|---|---|---|---|---|---|---|---|
| (1) | (2) | (3) | (4) | (5) | (6) | (7) | (8) | (9) | (10) |
| N4550 | Complex | IA | ⋯ | ⋯ | 177 | 0.62 | ⋯ | 0.74 | 5.8 |
| N4551 | ⋯ | ⋯ | ⋯ | ⋯ | 69 | 0.31 | ⋯ | ⋯ | ⋯ |
| N4552 | Patch[e] | IA | ⋯ | ⋯ | 124 | 0.07 | ⋯ | 0.08 | 4.8 |
| N4564 | ? | ⋯ | ⋯ | ⋯ | 47 | 0.48 | ⋯ | ⋯ | ⋯ |
| N4570 | ⋯ | ⋯ | ⋯ | ⋯ | 158 | 0.59 | ⋯ | ⋯ | ⋯ |
| N4589 | Lane and clumps | IS | 175 | 0.1 | 93 | 0.23 | −82 | 1.87 | 6.0 |
| N4621 | ? | ⋯ | ⋯ | ⋯ | 163 | 0.38 | ⋯ | ⋯ | ⋯ |
| N4623 | ? | ⋯ | ⋯ | ⋯ | 177 | 0.35 | ⋯ | ⋯ | ⋯ |
| N4636 | Irregular lane | IA | 109 | 0.2 | 151 | 0.02 | 42 | 0.34 | 5.1 |
| N4649 | ? | ⋯ | ⋯ | ⋯ | 100 | 0.12 | ⋯ | ⋯ | ⋯ |
| N4697 | Disk | RS | 65 | 0.2 | 64 | 0.38 | −1 | 0.33 | 5.5 |
| N4742 | ? | ⋯ | ⋯ | ⋯ | 74 | 0.43 | ⋯ | ⋯ | ⋯ |
| N4874 | ⋯ | ⋯ | ⋯ | ⋯ | 29 | 0.06 | ⋯ | ⋯ | ⋯ |
| N5102 | Curved wisp | RA | ⋯ | ⋯ | 48 | 0.30 | ⋯ | 0.12 | 4.0 |
| N5322 | Disk | RS | 96 | 0.1 | 93 | 0.35 | −3 | 0.34 | 5.9 |
| N5813 | Lane | RS | 150 | 0.1 | 141 | 0.10 | −9 | 0.06 | 5.0 |
| N5845 | Lane | RS | 144 | 0.1 | 145 | 0.27 | 1 | 0.13 | 4.4 |
| N5982 | ⋯ | ⋯ | ⋯ | ⋯ | 107 | 0.38 | ⋯ | ⋯ | ⋯ |
| N6166 | Sinusoidal lane | IA | 78 | 0.3 | 28 | 0.14 | −50 | 2.65 | 7.2 |
| IC1459 | Warped lane | RA | 141 | 0.2 | 128 | 0.26 | −13 | 0.13 | 5.3 |
| N7252 | ⋯ | ⋯ | ⋯ | ⋯ | 93 | 0.06 | ⋯ | ⋯ | ⋯ |
| N7332 | ? | ⋯ | ⋯ | ⋯ | 159 | 0.48 | ⋯ | ⋯ | ⋯ |
| N7457 | ⋯ | ⋯ | ⋯ | ⋯ | 130 | 0.38 | ⋯ | ⋯ | ⋯ |
| N7626 | Lane | RA | 171 | 0.2 | 6 | 0.10 | 15 | 0.11 | 4.3 |
| N7768 | Ring | RS | 84 | 0.7 | 61 | 0.26 | −23 | 0.36 | 6.0 |

Table 3. Basic data and X-ray masses of sample members.

| Ident | Type | $d$ (Mpc) | $M_B$ | $\log M_{\rm X}$ ($M_\odot$) | |
|---|---|---|---|---|---|
| (1) | (2) | (3) | (4) | (5) | |
| N221 | CE2 | 0.7 | −15.34 | ··· | |
| N524 | LAT | 48.2 | −22.04 | 9.71 | |
| N596 | E2P | 39.6 | −21.33 | < 8.87 | |
| N720 | E5 | 35.7 | −21.61 | 9.49 | |
| N1023 | LBT− | 15.8 | −21.04 | ··· | |
| N1052 | E4 | 29.3 | −21.05 | ··· | |
| N1172 | E1∗ | 47.1 | −20.81 | < 8.12 | |
| N1275 | P | 105 | −22.47 | ··· | |
| N1316 | PLXS0P | 25.4 | −22.35 | 10.03 | |
| N1399 | E1P | 26.4 | −21.56 | 9.68 | |
| N1400 | LA− | 31.3 | −20.86 | < 6.82 | |
| N1426 | E4 | 30.9 | −20.20 | ··· | |
| N1439 | E1 | 30.9 | −20.38 | ··· | |
| N1600 | E3 | 98.6 | −23.18 | 10.52 | |
| N1700 | E4 | 81.1 | −22.29 | ··· | |
| N2110 | SAB0− | 45.7 | −19.30 | ··· | a |
| N2832 | E+2∗ | 136 | −22.88 | 10.97 | |
| N3115 | L−/ | 9.6 | −19.97 | < 7.95 | |
| N3311 | LA0∗ | 68.1 | −23.02 | ··· | a |
| N3377 | E5+ | 13.3 | −19.49 | < 7.50 | |
| N3379 | E1 | 13.2 | −20.17 | < 8.95 | |
| N3384 | LBS−∗ | 12.2 | −19.61 | < 7.93 | |
| N3599 | LA∗ | 20.4 | −18.86 | ··· | a |
| N3605 | E4+ | 20.4 | −18.31 | 6.58 | |
| N3608 | E2+ | 20.4 | −19.87 | ··· | |
| N4150 | LAR0$ | 14.6 | −18.36 | ··· | |
| N4168 | E2 | 43.8 | −21.26 | 8.10 | |
| N4261 | E2+ | 41.0 | −21.74 | 9.70 | |
| N4278 | E1+ | 14.6 | −19.80 | ··· | |
| N4365 | E3 | 20.2 | −20.89 | 8.82 | |
| N4374 | E1 | 20.7 | −21.45 | 9.41 | |
| N4387 | E5 | 20.7 | −18.71 | ··· | a |
| N4434 | E0 | 20.3 | −18.71 | ··· | |
| N4472 | E2 | 20.3 | −22.21 | 10.21 | |
| N4473 | E5 | 20.8 | −20.38 | < 8.62 | |
| N4476 | LAR−∗ | 20.7 | −18.57 | < 7.73 | |
| N4478 | E2 | 20.7 | −19.44 | < 8.25 | |
| N4486 | E+0+P | 20.7 | −22.06 | ··· | |
| N4494 | E1+ | 22.4 | −21.07 | ··· | |

Table 3—Continued

| Ident | Type | $d$ (Mpc) | $M_B$ | $\log M_{\rm X}$ ($M_\odot$) | |
|---|---|---|---|---|---|
| (1) | (2) | (3) | (4) | (5) | |
| N4550 | LB+$/ | 25.2 | −19.70 | < 7.92 | |
| N4551 | E3∗ | 20.7 | −18.86 | ⋯ | a |
| N4552 | E0 | 20.8 | −20.75 | 8.99 | |
| N4564 | E6 | 20.7 | −19.62 | < 8.15 | |
| N4570 | L/ | 25.2 | −20.31 | ⋯ | |
| N4589 | E2 | 36.6 | −21.23 | 9.20 | |
| N4621 | E5 | 20.7 | −20.93 | 8.76 | |
| N4623 | LB+$/ | 25.2 | −18.71 | ⋯ | |
| N4636 | E0+ | 19.9 | −21.30 | 9.17 | |
| N4649 | ⋯ | 20.7 | −21.81 | 9.80 | |
| N4697 | E6 | 20.4 | −21.52 | 9.14 | |
| N4742 | E4∗ | 22.3 | −19.71 | ⋯ | |
| N4874 | E+0 | 138 | −23.39 | ⋯ | a |
| N5102 | ⋯ | 3.3 | −17.03 | < 7.05 | |
| N5322 | E3+ | 42.1 | −22.03 | < 9.46 | |
| N5813 | E1+ | 31.9 | −21.13 | ⋯ | |
| N5845 | E3 | 31.9 | −19.17 | ⋯ | a |
| N5982 | E3 | 59.3 | −21.89 | < 9.78 | |
| N6166 | E+2P | 182 | −23.54 | ⋯ | a |
| IC1459 | E3+ | 33.1 | −21.72 | 9.56 | |
| N7252 | RLAR0∗ | 93.8 | −22.80 | ⋯ | a |
| N7332 | LP/ | 27.3 | −20.46 | < 8.81 | |
| N7457 | LAT−$ | 18.5 | −20.06 | ⋯ | |
| N7626 | E1P∗ | 74.3 | −22.36 | 9.95 | |
| N7768 | E2 | 165 | −23.11 | ⋯ | a |

Table 4. Sample members with extended radio structure

| Object | Description of radio structure | PA mas | PA arcsec | Ref. |
|---|---|---|---|---|
| NGC 1052 | clear jet, curvature mas → arcsec scales | $\sim 60°$ | $95° \pm 3°$ | 1, 2 |
| NGC 1275 | knotty jet | $195°$ | $160°$ | 3, 4 |
| NGC 1316 | jet + lobes | | $126° \pm 14°$ | 5 |
| NGC 1399 | low luminosity jet | | $\sim -20°$ | 6 |
| NGC 1600 | double structure, jet not detected by VLA | | $98° \pm 2°$ | 5 |
| NGC 2110 | curved jet | | $0° : \sim 30°$ | 7 |
| NGC 4261 | straight, well defined jet | $3°$ | $88° \pm 1°$ | 8, 5 |
| NGC 4278 | marcsec extended, large scale H I cloud | $-28° \pm 10°$ | $-40° \pm 3°$ | 1 |
| NGC 4374 | marcsec extended, one sided extension $\sim 10''$ | $10°$ | $0°$ | 8, 9 |
| NGC 4472 | $\sim 3$ arcsec jet | | $83° \pm 4°$ | 5 |
| NGC 4486 | well defined, one-sided jet | | $290°$ | 10 |
| NGC 4552 | mas extended, optical jet | $35°$ | | 8, 11 |
| NGC 4636 | jetlike, S–shaped | | $33° \pm 8°$ | 5 |
| NGC 4874 | small WAT source, possible jets | | $\sim 50°$ | 12 |
| NGC 5322 | broad jet, low total power | | $177°$ | 13 |
| NGC 6166 | clear jet on mas scales, knots on arcsec scales | $90°$ | $\sim 90°$ | 14 |
| NGC 7626 | well defined jet | | $35° \pm 5°$ | 15, 5 |

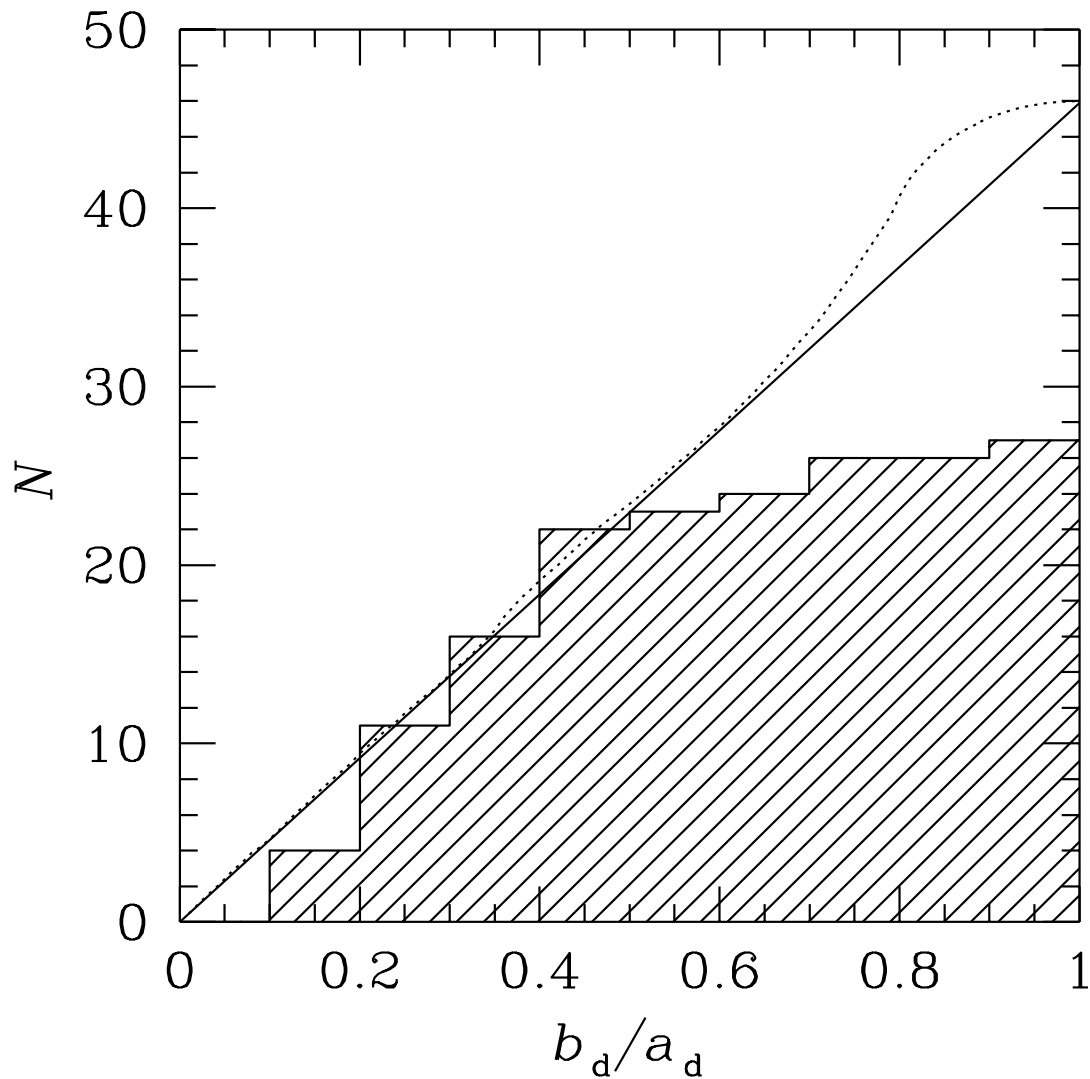

Fig. 1.— The cumulative distribution of $\frac{b_d}{a_d}$, the observed axis ratio of the dust feature. The solid line shows the 'true' distribution, assuming circular orbits and completeness for $\frac{b_d}{a_d} < 0.5$. There seems to be a bias against the detection of face-on dust features. Accounting for this bias, we estimate $78 \pm 16\,\%$ of early-type galaxies have a dusty core. The dotted line shows the expected distribution for dust on elliptical orbits, with intrinsic axis ratio 0.8. The estimate for the total number of dusty cores is not dependent on the assumption that the dust is on circular orbits.

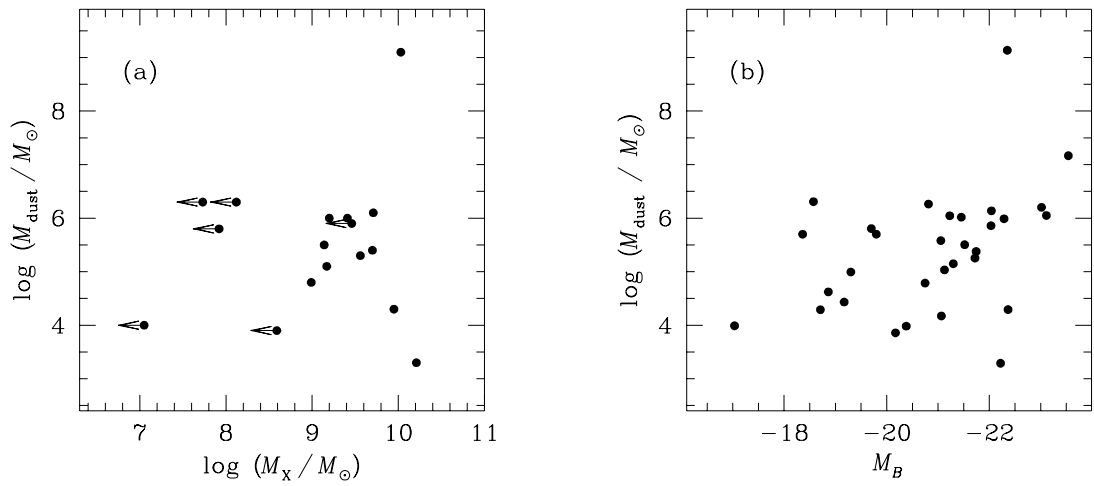

Fig. 2.— Scatter plots of nuclear dust mass vs. X-ray mass (a) and absolute magnitude (b). Arrows indicate upper limits. The dust mass does not correlate with either the X-ray mass or the absolute magnitude of the galaxy.

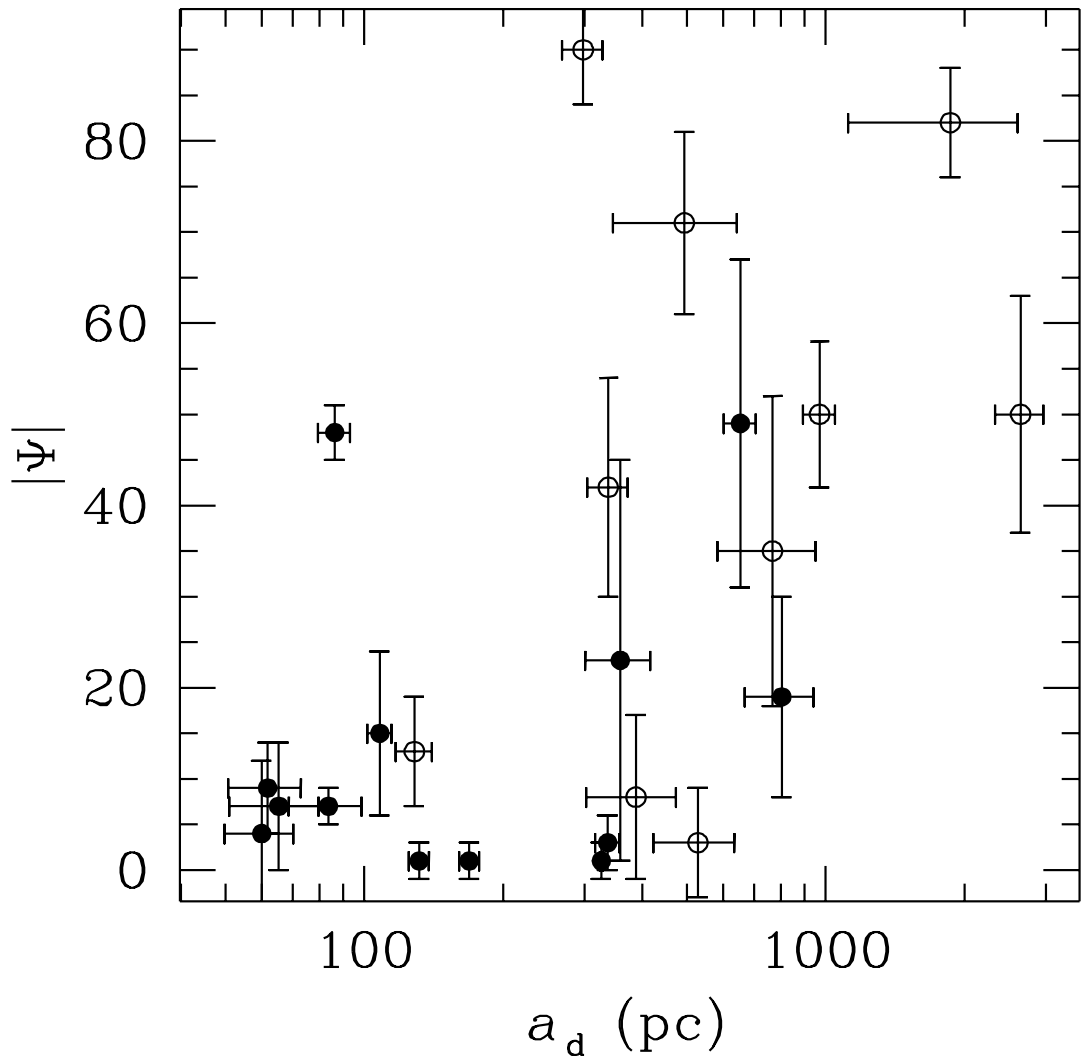

Fig. 3.— Misalignment angle $|\Psi|$ vs. semi-major axis size $a_d$ of the dust feature. Filled circles represent regular dust, open circles irregular dust. Errors are estimated from the residual images of the ellipse fits to the galaxies (Fig. 8). The misalignment angle is distributed randomly for galaxies with $a_d > 250\,\mathrm{pc}$, but the dust is rather well aligned for galaxies with $a_d < 250\,\mathrm{pc}$.

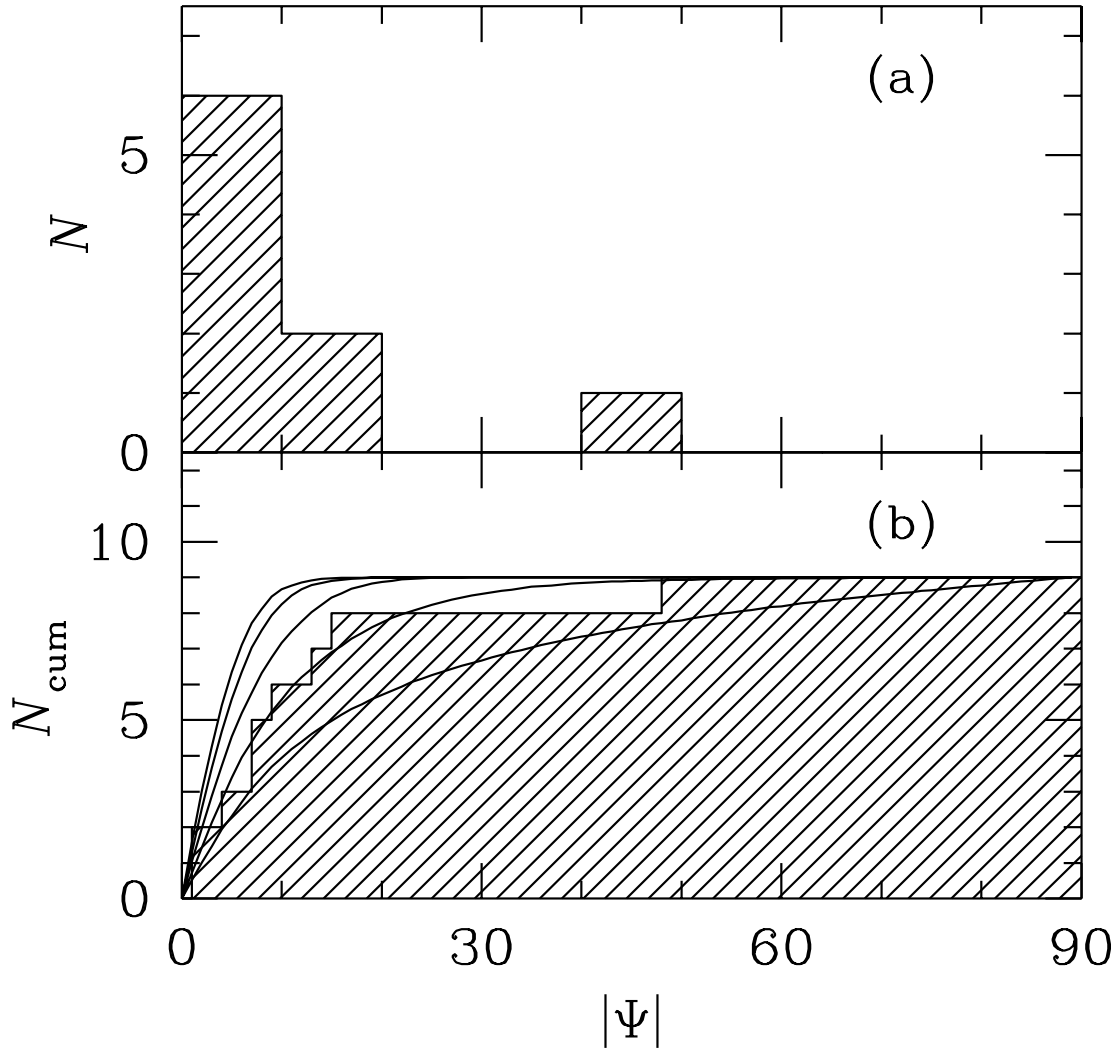

Fig. 4.— Distribution of $|\Psi|$ for dust features with $a_d < 250\,\text{pc}$ (a). In (b), the cumulative distribution is shown. Lines are of constant triaxiality parameter $T$, assuming intrinsic short axis rotation, and a bias against the detection of dust features with axis ratios $\frac{b_d}{a_d} > 0.5$. A measurement uncertainty of $5°$ was accounted for. From top to bottom $T = 0.2, 0.4, 0.6, 0.8$ and $1$ (prolate). An average triaxiality $\langle T \rangle < 0.4$ can be ruled out at the 95 % confidence level. The best fit is obtained for $\langle T \rangle = 0.8$.

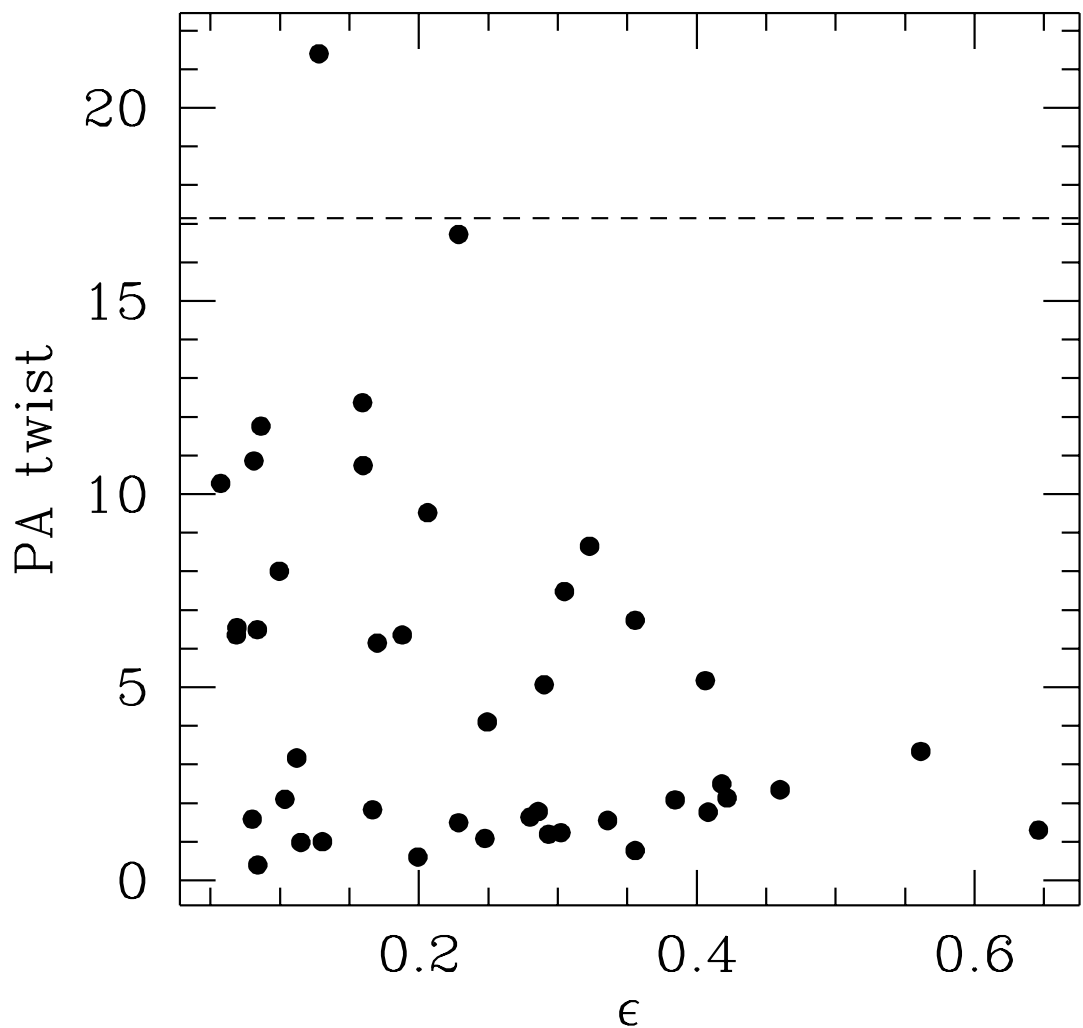

Fig. 5.— Position angle twist of the stellar component between $r = 25''$ and $r = 1\rlap{.}''5$, as a function of $\epsilon$. The rms position angle twist equals $7°$. The dashed line indicates the expected $1\sigma$ spread of $17°$, for a change in triaxiality $\Delta T = 0.4$. The data exclude this value of $\Delta T$. A Student $t$ test gives an upper limit of $\Delta T = 0.20$, at the 95 % confidence level.

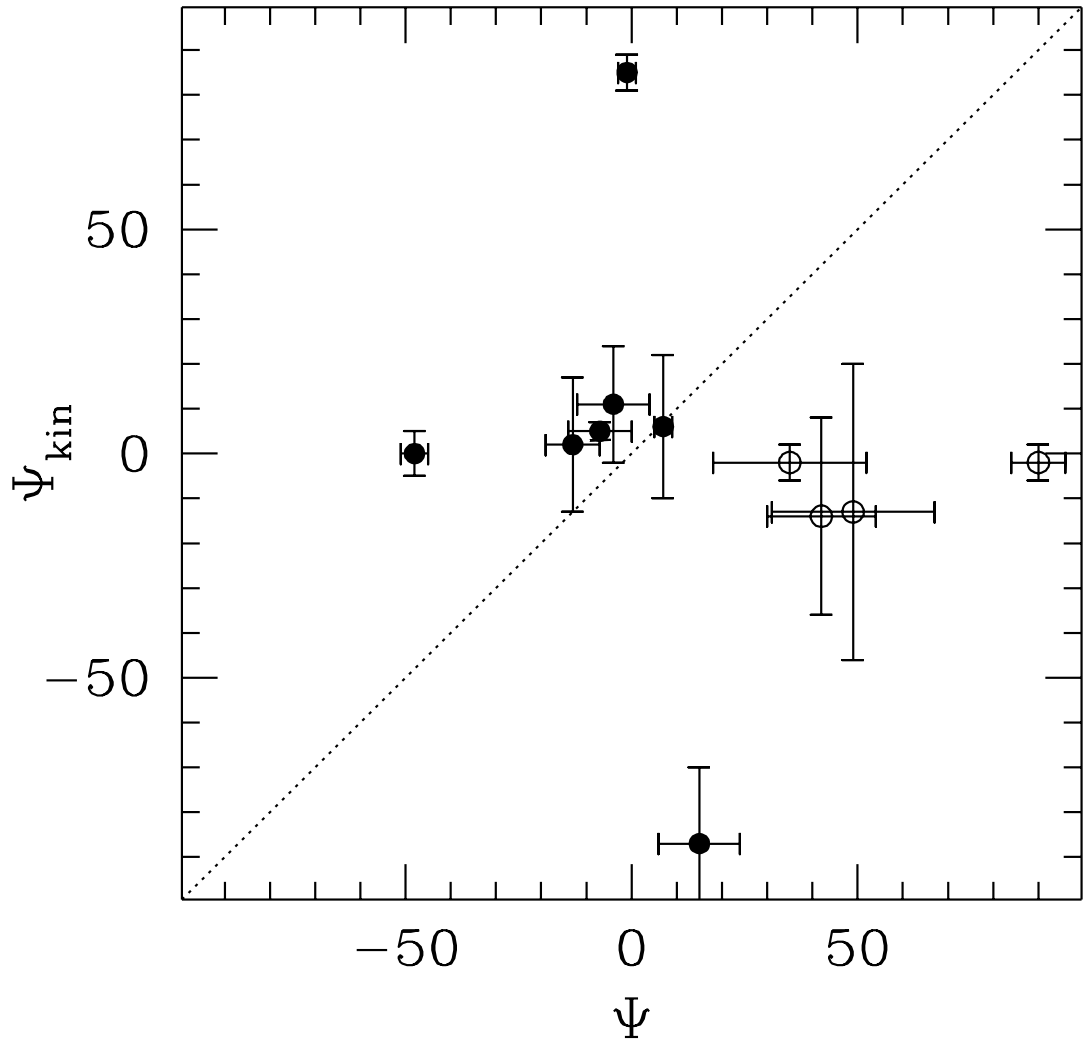

Fig. 6.— Kinematic misalignment $\Psi_{\rm kin}$ of the stars and misalignment of dust major axis and galaxy major axis $\Psi$. Filled circles represent galaxies with $a_{\rm d} < r_{\rm rel}$. If $\phi_{\rm int} = 0$ and dust lanes are relaxed, the points are expected to follow the diagonal line $\Psi_{\rm kin} = \Psi$. Since no galaxies with misalignments significantly different from zero have $\Psi_{\rm kin} = \Psi$ the dust and the stars do not seem to be kinematically coupled. This suggests an external origin of the dust.

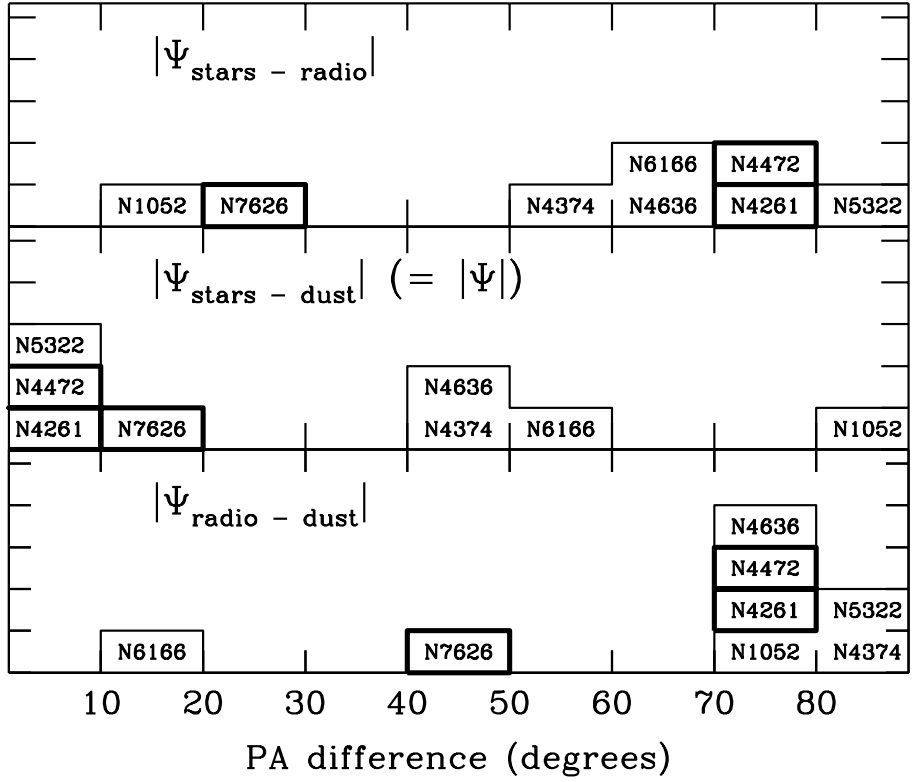

Fig. 7.— Distribution of $|\Psi_{stars-radio}|$, $|\Psi|$ and $|\Psi_{radio-dust}|$ for all sample members with double radio structure and a dust lane. Galaxies with dust lanes that have $a_d < r_{rel}$ are indicated by thick boxes. We confirm the conclusion of Kontanyi & Ekers (1979) that $|\Psi_{radio-dust}|$ peaks near 90°.

Fig. 8.— Mosaic of all cores which show clear evidence for dust absorption. The ellipse indicates the shape of the galaxy. The line in the bottom right corner of each panel shows the estimated position angle of the dust feature. The scale of the images is $5\rlap{.}''5 \times 5\rlap{.}''5$. Many dust features have an irregular appearance. In several cases, the dust is not aligned with the major axis of the galaxy.